\documentclass[superscriptaddress,groupedaddress,pre,twocolumn]{revtex4}  

\usepackage{graphicx}  
\usepackage{dcolumn}   
\usepackage{bm}        
\usepackage{amssymb}   
\usepackage{color}
\usepackage{setspace}
\usepackage{amsmath}
\usepackage{ulem}
\usepackage{epstopdf}

\begin{document}

\title{Predicting 2D Turbulence}

\author{R.T. Cerbus}
\email{rtc17@pitt.edu}
\affiliation{Department of Physics and Astronomy, University of Pittsburgh, 3941 O'Hara Street, Pittsburgh PA 15260}
\affiliation{Fluid Mechanics Unit, Okinawa Institute of Science and Technology Graduate University, Okinawa 904-0495, Japan}
\author{W.I. Goldburg}
\affiliation{Department of Physics and Astronomy, University of Pittsburgh, 3941 O'Hara Street, Pittsburgh PA 15260}

\begin{abstract}

Prediction is a fundamental objective of science. It is more difficult for chaotic and complex systems like turbulence. Here we use information theory to quantify spatial prediction using experimental data from a turbulent soap film. At high Reynolds number $Re$ where a cascade exists, turbulence is becoming easier to predict as the inertial range broadens. A transition corresponding to the emergence of a cascade  at low $Re$ is detected by looking at turbulence through prediction.

\end{abstract}

\maketitle

\section{Introduction}

According to many  textbooks, a hallmark of turbulence is its unpredictability \cite{tritton1988,tennekes1972}. Here we address this issue using experimental data from a turbulent soap film. The starting point is Shannon's information theory \cite{shannon1964,cover1991,brillouin1962}, where in Neil Gershenfeld's words, ``...information is what you don't already know" \cite{gershenfeld2000}. Our experiment conveys information about the physical state of the system. The amount of previously unknowable information is our measure of unpredictability. 

Our objective is to quantify the prediction of turbulent velocity fluctuations and in the process characterize turbulence. We will measure both the limits on making predictions and how much we need to know to do so \cite{crutchfield2012}. This approach parallels the use of Lyapunov exponents to characterize the sensitivity to initial conditions (unpredictability) of chaotic systems \cite{baker1996}. The main finding is a transition in our ability to predict, corresponding to the emergence of a cascade.

The turbulent cascade envisioned by Richardson and described mathematically by Kolmogorov is the prevalent picture of turbulence \cite{davidson2004}. In this picture, energy (or enstrophy in two dimensions) is transported across scales from some injection scale until it reaches a dissipative scale and the cascade ends. This cascade exists in both three dimensional (3D) and two dimensional (2D) turbulence, such as the one studied here. We argue that a cascade influences prediction, as discussed below.

The central quantity in information theory is the entropy density $h$ \cite{cover1991}. It is the information we receive per measurement (which in this case means a single velocity value). Think of this as the number of yes/no questions needed on average to determine the next measurement (not necessarily an integer) \cite{shannon1964}. This number can be reduced if the data is not random and structureless \cite{cover1991}. By looking at all previous data, we reduce the unpredictability to $h$. Of course, knowing the value of $h$ does not tell us how to make a prediction, only the limits on our ability to do so.

While $h$ is the amount of information that we don't know, we could also ask how much we do know. This is the excess entropy $E$, which is the information about correlations in the system \cite{schurmann1996,crutchfield2003}. It is the reduction of unpredictability. Accurate prediction requires an amount of information at least equal to $E$ \cite{ellison2009}. Although $E$ further characterizes our ability to predict, we still must decide how to do so.

\begin{figure}[h!]
\centering
\includegraphics[scale = 0.33]{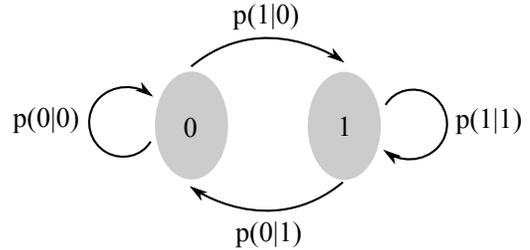}
\caption{A schematic of a binary system's states and transition probabilities between them (``digraph" \cite{ellison2009}). The transition probabilities are the conditional probabilities $p(0|1)$, $etc.$ In this notation, $p(0|1)$ means the probability of 0 given 1.}
\label{digraph2}
\end{figure}

Now we must decide how to make a prediction. There are many options, but our choice is to make a statistical model with a set of states and the probabilities to transition between them. For a binary system with $0s$ and $1s$, this will look like the schematic in Fig. \ref{digraph2}. There is more than one way to define which states to use and potential benefits from choosing them cleverly.

Starting with the states that are present in the data $U$, that set is then reduced by combining those states which statistically lead to the same future \cite{crutchfield2012}. This makes the connection with prediction clear. The information contained in these optimally predictive ``causal states" $S$ is the statistical complexity $C$ of Crutchfield \cite{shalizi2001,crutchfield2012,ellison2009}. It is defined so that it is zero for laminar flows and zero also for completely random velocity fluctuations. In both limits the system's prior behavior tells one nothing about velocity fluctuations to follow. It is known that $C \ge E$, but the reasons why are not always clear \cite{crutchfield2012,ellison2009}. More details on $h$, $E$ and $C$ can be found in the Appendices \ref{appendix:data}-\ref{appendix:C}.

This study focuses on predicting the spatial variations of turbulence. A prediction in space means that given the velocity $u$  at a point $x$, one anticipates the velocity at some other point $r$ away. Prediction is normally associated with time \cite{aurell1996,leith1972}, but there are several reasons for considering the spatial alternative.

We know that the temporal and spatial features of turbulence are distinct. The fundamental work of Kolmogorov dealt only with the spatial structure of turbulence \cite{kolmogorov1941,davidson2004}. Kraichnan and others have also shown that many of the essential features of turbulence are retained if one throws away temporal correlations but keeps spatial ones \cite{kraichnan1994, shraiman2000, falkovich2001}. Thus, a treatment of spatial prediction is arguably of more fundamental interest than temporal prediction, at least for turbulence.

For a specific application, consider airplane flight. The typical cruise speed of a Boeing 747 is $V \simeq 250$ m/s \cite{boeing}. Contrast this with the rms velocity fluctuations $\sigma$ of ``strong" atmospheric turbulence $\sigma \simeq 7$ m/s \cite{mcminn1997}. Since $\sigma/V \simeq 0.03$ is small, one must use Taylor's frozen turbulence hypothesis when discussing the turbulence the airplane encounters \cite{kellay2002,tennekes1972}. In other words, an airplane flies fast enough to sample only the spatial variations of turbulence. There is not enough time for the turbulent velocity field to evolve temporally.

We have previously used $h$ to characterize two-dimensional (2D) turbulence in a flowing soap film as a function of Reynolds number $Re$ \cite{cerbus2013}. Here we use $C$ and $E$ to go beyond this and fully characterize the prediction of turbulent velocity fluctuations. This leads to the following conclusions. (1) The presence of correlated velocity fluctuations $reduces$ the amount of information $C$ needed to predict. However, those same velocity correlations increase $E$. Thus, $C$ and $E$ may be used as an indicator of the presence of a turbulent cascade. (2) 2D turbulence becomes increasingly easy to predict as $Re$ increases. While this study is on an experimental 2D system, the arguments apply for 3D turbulence as well. Moreover, no specific assumptions about the data are made. Thus, this study serves as an experimental test bed for these tools, which can be used generally for other complex systems.

\section{Example}
\label{sec:example}

As a simple illustration of these ideas, consider a coin flipping experiment where each subsequent flip  will be the same as the previous one with probability $P \in [0,1]$ \cite{quax2013}. This is the statistical model for, $e.g.$, correlated random walks \cite{codling2008}. The statistical evolution of this system will look like Fig. \ref{digraph2} but with, $e.g.$, $p(0|0) = P$. 

If $P = 0.5$ we have the usual fair coin toss experiment, with $h = 1$ and $C = E = 0$, since this system is maximally uncertain but statistically simple to predict with no information being shared between the past and future. In this fully random case ($P = 0.5$) both 0 and 1 predict the same future, so they are combined into a single causal state. Of course, with only one causal state, $C = 0$ automatically (see Eq. \ref{eq:C}). 

\begin{figure}[h!]
\centering
\includegraphics[scale = 0.33]{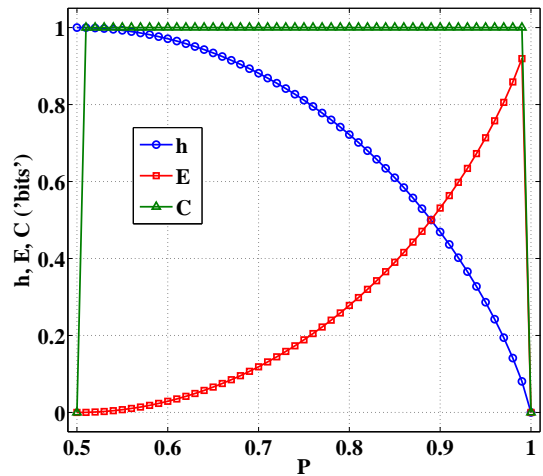}
\caption{Plot of the fundamental quantities $h$ $(\bigcirc)$, $E$ $(\square)$ and $C$ $(\triangle)$ for the simple example given here. Although $h$ and $E$ are continuous functions of $P$, $C$ is not.}
\label{simple_example}
\end{figure}

Consider now a slight deviation of $P$ from 0.5. Now $C = 1$ since we will always need to know 1 bit of information (the previous flip) to predict the future. We can also calculate $h$ and $E$ (see Appendices \ref{appendix:h} and \ref{appendix:E}), which are plotted together with $C$ vs. $P$ in Fig. \ref{simple_example}. Since $P > 0.5$ means more predictable, it is clear that $h$ should decrease with increasing $P$, while $E$ should increase.

This example highlights the difference between $E$ and $C$, the crypticity $\chi \equiv C - E$ \cite{ellison2009,mahoney2011}. Here $C = E + h$, which is a unique feature of this system being first-order Markovian \cite{crutchfield2003}. The extra information needed to predict beyond $E$ is due to the randomness still intrinsic in the causal states themselves. There are many examples for which $C \ne E$ \cite{crutchfield2009,ellison2009}, but this is not always so.

An important lesson we learn from this example is that $h$, $E$ and $C$ were all necessary to characterize this system's behavior. For $P$ only slightly different from 0.5, $h$ and $E$ will still suggest a nearly random system, much like a slightly biased coin. The fact that $C$ is large and not 0 (its random value), shows that there are important correlations not present in a simple biased coin system. The system is both unpredictable (large $h$) and difficult to predict (large $C$). A similar result will be found for the low Reynolds number flow in Sec. \ref{sec:results}.

\section{Experimental setup}

\begin{figure}[h!]
\hspace{-1.5em}
\includegraphics[scale = 0.19]{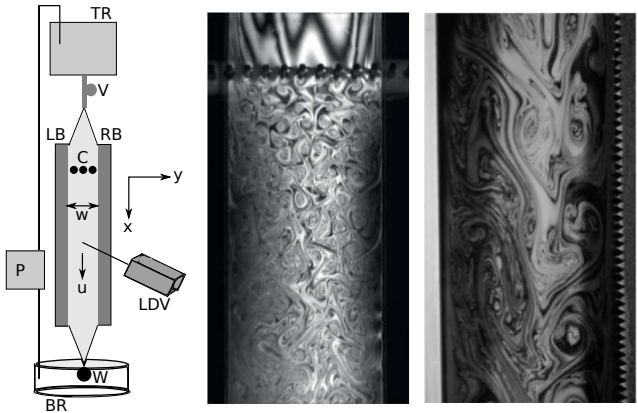}
\caption{Left: Experimental setup showing the reservoirs ($TR$, $BR$), pump ($P$), valve ($V$), comb ($C$), blades ($LB$, $RB$), LDV and weight ($W$). Middle: Fluctuations in film thickness from turbulent velocity fluctuations with smooth walls and a comb. Right: Thickness fluctuations with smooth and rough walls.}
\label{setup}
\end{figure}

Now consider a turbulent soap film, which is a good approximation to 2D turbulence since the film is only several $\mu$m thick \cite{kellay2002,boffetta2012}. The soap solution is a mixture of Dawn (2$\%$) detergent soap and water with 4 $\mu$m particles added for laser doppler velocimetry (LDV) measurements. Figure \ref{setup} contains a diagram of the experimental setup as well as thickness fluctuations visualized through thin film interference using a monochromatic light source. The thickness fluctuations act as a surrogate for velocity fluctuations \cite{kellay2002,boffetta2012}.

The soap film is suspended between two vertical blades. Nylon fishing wire connects the blades to the nozzle above and the weight below. The nozzle is connected by tubes to a valve and a top reservoir which is constantly replenished by a pump that brings the spent soap solution back up to the top reservoir. The flow is gravity-driven. Typical centerline speeds $\overline{u}$ are several hundred cm/s with rms fluctuations $u'$ ranging roughly from 1 to 30 cm/s. The channel width $w$ is usually several cm. The Reynolds number $Re = u'w/\nu$, where $\nu$ is the kinematic viscosity, thus ranges from 10 to 10,000.

Turbulence is generated using several different protocols. We can (1) insert a row of rods (comb) perpendicular to the film, (2) replace on or both smooth walls with rough walls (saw blades) with the comb removed and possibly a rod inserted near the top \cite{kellay2012}, or (3) use a comb with smooth walls as in (1) but now very near the top of the soap film where the flow is still quite slow. The comb teeth are $\sim 1$ mm in diameter and several mm apart. The saw blade teeth are $\sim 2$ mm tall and wide. 

When protocol (1) is used we almost always observe the direct enstrophy cascade \cite{kellay2002,boffetta2012}. If procedure (2) is used, we can observe an inverse energy cascade \cite{kellay2002,boffetta2012,kellay2012}, although this depends sensitively on the flux and $w$. When protocol (3) is used, we see no cascade at all.

\begin{figure}[h!]
\hspace{-1.5em}
\includegraphics[scale = 0.36]{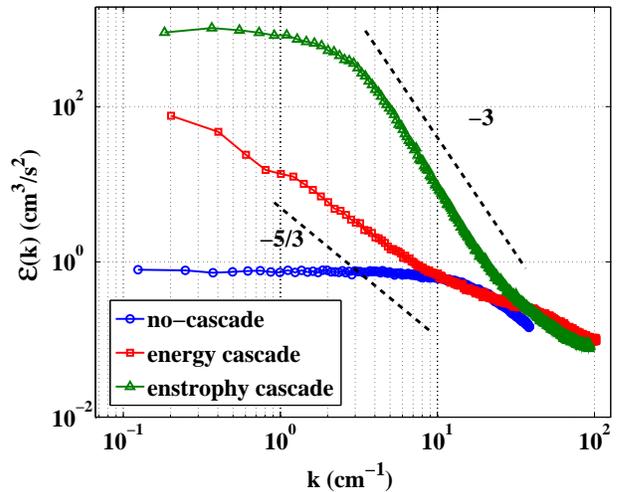}
\caption{Representative one-dimensional energy spectra in a log-log plot of $\mathcal{E}(k)$ vs. $k$. The enstrophy cascade ($\triangle$) has a slope close to -3 while the energy cascade ($\square$) has a slope close to -5/3. The flat curve ($\bigcirc$) has no cascade.}
\label{spectra}
\end{figure}

The type of cascade is identified by calculating the one-dimensional velocity energy spectrum $\mathcal{E}(k)$, where $\frac{1}{2}u'^2 = \int_0^{\infty} \mathcal{E}(k) dk.$ For the enstrophy cascade, $\mathcal{E}(k) \propto k^{-3}$ and for the energy cascade $\mathcal{E}(k) \propto k^{-5/3}$ \cite{kellay2002,boffetta2012}. A number of measurements were taken above the blades where the flow is slower. For protocol (3),  $\mathcal{E}(k)$ is flat and so apparently there is no cascade, although the flow is not laminar ($u' \neq 0$). See Fig. \ref{spectra} for some representative spectra. In Fig. \ref{Re_E&C} the data for $Re < 100$ have a flat $\mathcal{E}(k)$.

In all cases, we measure the longitudinal (streamwise) velocity component at the horizontal center of the channel. The data rate is $\simeq$ 5000 Hz and the time series typically had more than $10^6$ data points. For this system the time series is really a spatial series by virtue of Taylor's frozen turbulence hypothesis \cite{davidson2004,kellay2002,boffetta2012,tennekes1972}. This means that the spatial variations are swept through the LDV's measuring point by the mean flow so quickly that it is as if the LDV were scanning a frozen-in-time velocity field. This distinction between spatial and temporal is essential, as discussed above and in Ref. \cite{cerbus2013}.

\section{Results}
\label{sec:results}

\begin{figure}[h!]
\includegraphics[scale = 0.35]{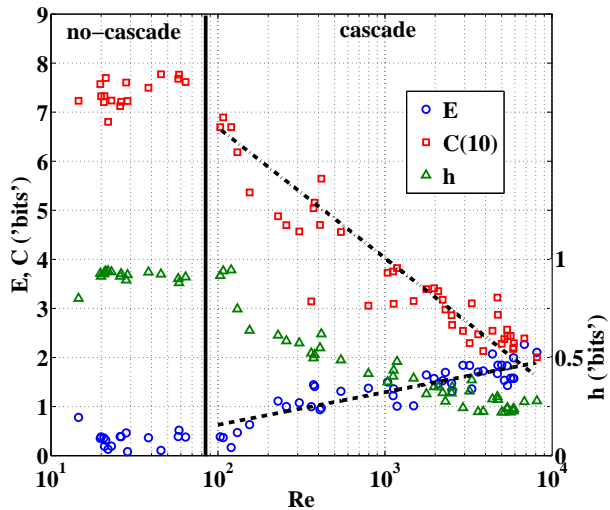}
\caption{The statistical complexity $C$ ($\square$), excess entropy $E$ ($\bigcirc$) and entropy density $h$ ($\triangle$) as functions of $Re$ for binarized ($A = 2$) data (see Appendix \ref{appendix:data} for details on binning). We plot $h$ on a different scale for better visibility. The maximum value of $h$ here is $\log_2 2 = 1$, which the no-cascade data for $Re < 100$ approach very closely. Here $L = 10$ and we used our MATLAB program with the $\chi^2$ test to calculate $C$ (see Appendix \ref{appendix:C} for details). The lines are not fits to the data but are meant to suggest the behavior of $C$ and $E$ as functions of $Re$. For the cascade region, $C$ and $h$ are decreasing functions of $Re$ while $E$ increases. The vertical line separates the data according to whether there is a cascade or not.}
\label{Re_E&C}
\end{figure}

The quantities $C$, $E$ and $h$ are plotted vs. $Re$ in Fig. \ref{Re_E&C}. The data are roughly divided in $Re$ into no-cascade (flat $\mathcal{E}(k)$ for $Re < 100$) and cascade (power law $\mathcal{E}(k)$ for $Re > 100$) regimes. Although $C$ and $E$ intersect at finite $Re \simeq 7000$ in Fig. \ref{Re_E&C}, this meeting point depends on the data analysis. In order to calculate probabilities from continuous data, one must bin the measurements. For different binning protocols we find a different meeting point. However, the $Re$-dependent behavior of $h$, $E$ and $C$ discussed below is the same. See Appendices \ref{appendix:data} and \ref{appendix:C} for more details on the treatment of the data.

\subsection{Cascade Turbulence}

Now consider the behavior of $h$, $E$ and $C$ in the ``cascade regime" of Fig. \ref{Re_E&C}, $Re > 100$. At these values of $Re$, $\mathcal{E}(k)$ shows power law scaling as in Fig. \ref{spectra}. Both energy and enstrophy cascade data are present. We see from Fig. \ref{Re_E&C} that the unpredictability ($h$) is decreasing, the amount of information needed to predict ($C$) is also decreasing, while information about correlations ($E$) is increasing (all logarithmically). The opposite trend in $Re$ for $E$ and $C$ is noteworthy. It is surprising that the behavior of $h$, $E$ and $C$ for $Re > 100$ does not depend on which cascade is present, only on whether or not there is a cascade at all.

The increase of $E$ with $Re$ can be understood from the traditional view that as $Re$ increases, the ``inertial range" of correlated scales broadens \cite{davidson2004}. The increase in correlations across spatial scales is reflected by an increase in $E$. We can go further to suggest a connection between $E$ and the broadness of the inertial range. Dimensional arguments suggest that the turbulent degrees of freedom go as $N \propto Re$ for the enstrophy cascade and $N \propto Re^{3/2}$ for the inverse energy cascade. In the 3D energy cascade, $N \propto Re^{9/4}$ \cite{landau1987}. Thus the behavior $E \propto \log_2 Re$ in Fig. \ref{Re_E&C} indicates that $E$ is a logarithmic measure of the extent of the inertial range.

An interpretation of the behavior of $C$ is also suggested by the traditional picture of 2D turbulence \cite{kellay2002,boffetta2012}. As $Re$ grows, the inertial range broadens, and more of the velocity fluctuations come under the governance of the cascade. Thus, the randomness $h$ will decrease, and because the cascade's structure is dominating, our  prediction cost $C$ decreases. This is the result of the general principle that patterns help us to predict \cite{shalizi2001}. Here the pattern is the cascade's structure.

Although turbulence has traditionally been thought of as unpredictable \cite{tritton1988,tennekes1972}, with $h$, $E$ and $C$ we see that the spatial predictability of (2D) turbulence is increasing with $Re$ in its fullest sense: we can predict further and more easily. This is in stark contrast to turbulence's increasing temporal unpredictability with $Re$, at least as evidence by numerical work \cite{aurell1996,leith1972}. This reiterates the important difference between time and space in turbulence, which is of fundamental interest and practical importance (recall the airplane).

\subsection{Transition to Cascade Turbulence}

Next consider the region of Fig. \ref{Re_E&C} labeled ``no-cascade". The absence of a cascade is evidenced by a lack of power law scaling in $\mathcal{E}(k)$ as in Fig. \ref{spectra}. Here $h$, $E$ and $C$ are relatively constant with respect to $Re$. It is notable that $h$ is very near to the random (white noise) value of $\log_2 2 = 1$, which is nothing like laminar flow where $h = 0$. When a cascade emerges at $Re \simeq 100$, all three quantities begin to change noticeably. This change in behavior is decidedly different from the laminar to turbulent transition which only involves the onset of fluctuations \cite{tritton1988,landau1987}.

The fluctuations of pre-cascade turbulence are apparently difficult to predict ($C$ is large in Fig. \ref{Re_E&C}). Moreover, the wide separation between $E$ and $C$ is surprising. We emphasize that $C$, $E$ and $h$ have made a clear distinction between simply unsteady velocity fluctuations and cascade turbulence. It is natural that tools designed to quantify randomness and order should be able to detect this transition.

Simulations of 3D turbulence have shown that statistics of the velocity derivatives are gaussian (or sub-gaussian) up until a small value of the Reynolds number \cite{schumacher2007,schumacher2014}. Below this value of Reynolds number, there is a ``regime which is a complex time-dependent flow rather than a turbulent one." They observe a transition similar to the one described here. Their transition is evidenced primarily by non-gaussian velocity derivative statistics. Recall that nongaussian statistics are a general feature of fully developed turbulence \cite{sreenivasan1997}.

\begin{figure}[h!]
\includegraphics[scale = 0.36]{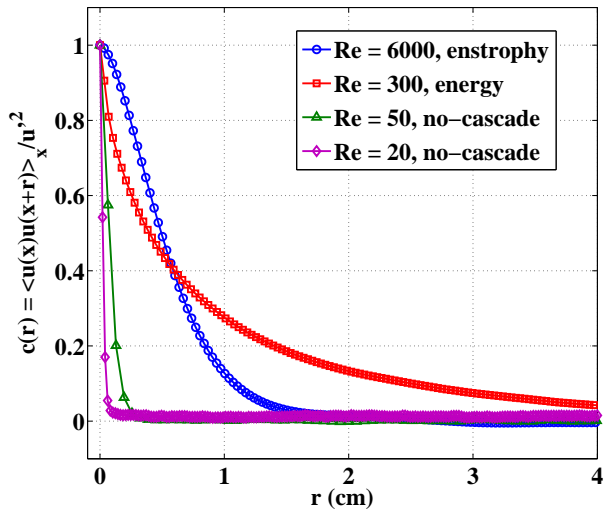}
\caption{The velocity autocorrelation function $c(r)$ plotted $vs.$ $r$ for several values of $Re$. For small $Re$, $c(r)$ quickly decays to zero, indicating little correlation in the velocity $u$. For larger $Re$, where Fig. \ref{spectra} indicates spatial structure, there is a wider range of correlated scales. The $Re = 300$ curve has a longer correlation length $L$ than the higher $Re = 6000$ curve presumably because this lower $Re$ curve corresponds to an inverse energy cascade. The inverse energy cascade is supposed to involve larger length scales than the enstrophy cascade \cite{kellay2002,boffetta2012}.}
\label{correlation}
\end{figure}

We also resort here to a more traditional tool from turbulence, the correlation function $c(r) \equiv \langle u(x)u(x+r) \rangle_x / u'^2$ plotted in Fig. \ref{correlation} \cite{davidson2004}. $c(r)$ has typically been thought of as a tool for determining the range of length scales over which $u$ is correlated. $c(r)$ is telling us that for small $Re \le 100$, the range of scales over which $u$ is correlated is very small.

Figures \ref{spectra} and \ref{correlation} both indicate that for $Re \le 100$ the data is like white noise. The values of $h \simeq 1$ and $E \simeq 0$ in Fig. \ref{Re_E&C} reinforce this interpretation. On the other hand, if the fluctuations were truly like white noise, then $C$ should also be zero in this regime, which it is not. Recall that in the simple example from Sec. \ref{sec:example}, $C$ is large when $h$ and $E$ are close to their random values. The data are nearly random but have an explicit albeit short dependence on the past which drives $C$ from zero to its maximum value. If we were to only look at $h$ (or $E$), we would miss that there is nontrivial (non-random) behavior for low $Re$.

We have yet to understand why self-similar turbulence emerges from this ``complex, time-dependent flow" \cite{schumacher2007}. One sees from another nonlinear system, Rayleigh-Benard convection, that there is a lot to be learned at modest levels of excitation \cite{kadanoff2001}.

The traditional approaches to the laminar-turbulent transition deal with instabilities of the laminar flow \cite{tritton1988,brandstater1983}. Whether it is the quasi-periodicity of Landau \cite{landau1987} or the nonperiodicity of Ruelle and Takens \cite{ruelle1971}, none of these approaches deal with the development of a cascade \cite{swinney1978}. And yet a cascade is always present in ``fully-developed turbulence" \cite{kolmogorov1941,davidson2004}. How does this cascade emerge? New approaches and models are necessary to understand how cascade behavior develops out of a ``complex, time-dependent flow" \cite{schumacher2007}. Since this development is clearly visible in Fig. \ref{Re_E&C}, an information theory approach seems promising.

\begin{figure}[h!]
\includegraphics[scale = 0.35]{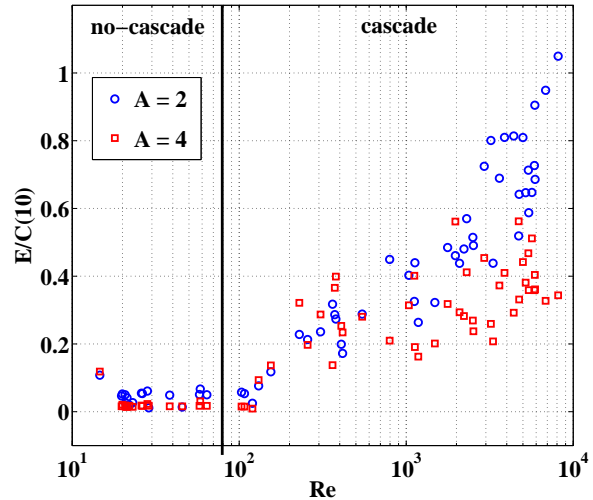}
\caption{The predictive efficiency $E/C$ plotted vs. $Re$ using the same data as in Fig. \ref{Re_E&C} as well as a quaternary partition $A = 4$ with partition walls placed symmetrically with respect to the mean (see Appendix \ref{appendix:data} for details on binning). We used $L = 10$ for both partitions (see Appendix \ref{appendix:C}). Here we find that $E/C$ is increasing only after a cascade develops.}
\label{pred_eff}
\end{figure}

We suggest an information-theoretic indicator of a cascade. Based on the above arguments, large $E$ and $1/C$ should both indicate a well-developed cascade. With that in mind, we can also consider the ``predictive efficiency" $E/C$ \cite{wiesner2012}, which is an increasing function of $Re$, as shown in Fig. \ref{pred_eff} for two different binning protocols. The ratio $E/C$ tells us the fraction of the information needed to predict $C$ that is due to correlations $E$. It is nearly zero when no cascade is present and grows smoothly after one has emerged. This shows that $E/C$ is a nice tool for studying the transition to cascade turbulence.

Besides this cascade transition, the laminar to fluctuation transition is also of interest. Unfortunately, we are not able to access a truly laminar regime with our apparatus. For laminar flow and this geometry, $h = E = C = 0$ \cite{crutchfield2012}. Looking at Fig. \ref{Re_E&C}, and with the reasonable assumption that $h$ and $C$ are continuous functions of $Re$, one expects a local maximum in $C$ and $h$ at some low value of $Re$. This maximum would correspond to a special transition in the evolution of the flow between laminar and turbulent behavior. The observation of this maximum requires a different experimental setup.

\section{Conclusion}

The approach here is not limited to incompressible Navier-Stoke's turbulence. In fact it is useful for any nonlinear system, even those for which one does not know the equations of motion. When we think of turbulence in terms of information and prediction, we can make new distinctions and draw new insights. We have been able to highlight a cascade transition and have seen that spatially, turbulence is becoming easier to predict statistically as $Re$ increases. As for our airplane, Figs. \ref{Re_E&C} and \ref{pred_eff} bring bittersweet news. Although its passengers will certainly experience a rougher flight as $Re$ increases, at least they won't be as surprised.

We would like to thank D. P. Feldman for explaining several concepts to us and for making his excellent lecture notes available online. C. J. Ellison was kind enough to explain some of the finer points of the formalism to us. We are also indebted to M. Bandi for providing numerous suggestions and insights. The criticisms and suggestions from several referees have also been beneficial. This work is supported by NSF Grant No. 1044105 and by the Okinawa Institute of Science and Technology (OIST). R.T.C. is also supported by a Mellon Fellowship through the University of Pittsburgh.

\begin{appendix}

\section{Data}
\label{appendix:data}

The approach used here is data driven. We are given a data stream and use it to say something about the system that made it. The main assumption is that the system is stationary \cite{cover1991,crutchfield2012}. We don't appeal to the Navier-Stoke's equation or any of Kolmogorov's universality assumptions \cite{davidson2004,kolmogorov1941}. This method is generally applicable to many types of systems.

The formalism is now introduced. In the discussion that follows an uppercase $U$ denotes the data (the random variable, the message) with possible velocity values $\mathcal{U}$ and the lowercase $u$ denotes a particular member of that set. We can also consider groups of length $L$ denoted by the set $\mathcal{U}^L$ and its particular members $u^L$. We are interested in treating a group because of the correlations that may exist between its members. Overhead arrows indicate a direction in the 1D data set relative to an arbitrary reference point $x$. For example, $\overrightarrow{U^L}$ refers to any block of data of size $L$ taken to the right of $x$. For example, if $L = 3$, then a particular block $\overrightarrow{u^3}$ is as below
\begin{equation*}
...u_{x-\Delta x},u_{x},\overrightarrow{u_{x+\Delta x},u_{x+2\Delta x},u_{x+3\Delta x}},u_{x+4\Delta x},...
\end{equation*}
where $\Delta x$ is the spatial resolution. If no $L$ is mentioned, the block is (semi-)infinite.

Let $U$ be a velocity component in the soap film, which is characterized by the experimental probability distribution $P(U)$. The focus is on the information shared between different directions $\overleftarrow{U}$ and $\overrightarrow{U}$ relative to the arbitrary point $x$ \cite{crutchfield1997,crutchfield2012}. If we had data with explicit time dependence, we would talk about the past, future and present \cite{crutchfield2012}.

In order to use this formalism with turbulence, the continuous experimental data must be converted to symbols \cite{daw2002}. A partition is defined which assigns data values in specific ranges to unique symbols \cite{daw2002, schurmann1996}. This is usually referred to as binning the data. All experiments of continuous systems do this because of limited resolution $\epsilon$. There are numerous previous studies where even binarizing a turbulent velocity signal has given more insight than traditional techniques \cite{daw2002,palmer2000,lehrman2001,cerbus2013}.

In this work we primarily use a binary partition (alphabet size $A = 2$) with the single partition wall located at the mean velocity. This smaller alphabet allows us to use a larger $L$ with confidence and so cover a wider range of length scales in our analysis. Just as with $h$ in Ref. \cite{cerbus2013}, we have found that the general behavior of $C$ and $E$ with respect to $Re$ is independent of the partition size; partitions of sizes $A = 4$, $8$ gave similar results. Here the choice was made to use the same alphabet size $A$ for all $Re$. This was done so that all data, if random, would have the same maximum value of $h = \log_2A$. Thus, all data are treated at the same level of description. Of course, there are alternative choices for setting the partition size.

\section{Entropy density $h$}
\label{appendix:h}

We have already spoken of the entropy density $h$ as a measure of unpredictability. The definition of entropy we are most familiar with is \cite{cover1991,shannon1964}
\begin{equation}
H(U) = -\sum_{u \in U} p(u) \log_2 p(u),
\end{equation}
with units of ``bits". This is the unpredictability of single data points given no immediate knowledge of any previous data points. An example of this would be estimating the unpredictability of letters in the English language based solely on the frequency of the letters and not on words. 

Consider two examples. First look at a random string of 1s and 0s where $p(0) = p(1) = 0.5$. Here $H = 1$ is the maximum possible value. Next consider a periodic string such as ``...0101...". Here again $p(0) = p(1) = 0.5$, and so here also $H = 1$. However, something is wrong since a periodic string should be perfectly predictable.

Since this definition of unpredictability misses any structure or correlations extending across scales, it is generalized to the block entropies \cite{schurmann1996,crutchfield2003}
\begin{equation}
H_L = H(U^L) = -\sum_{u^L \in U^L} p(u^L) \log_2 p(u^L).
\end{equation}
This is the unpredictability of blocks of data. Of course, if we want to go back to looking at the unpredictability of a single data point, we can manipulate the $H_L$. The unpredictability of a single data point knowing $L$ immediately previous data points is
\begin{equation}
h_L = H_{L+1} - H_L.
\label{eq:hL}
\end{equation}
The $L$-dependence is inconvenient, but if we make $L$ large enough $h_L$ will become $L$-independent (for most systems) \cite{schurmann1996,crutchfield2003}. We are now ready to introduce the entropy density
\begin{equation}
h = \lim_{L \rightarrow \infty} h_L = H(\overrightarrow{U^1} | \overleftarrow{U})
\label{eq:h}
\end{equation}
with an equivalent definition in terms of the conditional entropy \cite{cover1991}. This says explicitly how unpredictable a single data point is given all previous ones.

To further develop intuition for how $h$ is associated with unpredictability, recall the Lyapunov exponents \cite{baker1996}. If a system is chaotic, its largest Lyapunov exponent $\lambda$ is greater than 0 \cite{baker1996}. If our measurement has a resolution of $\epsilon$ and we enforce a tolerance of $\Delta$, then our system is typically predictable up to a distance of $\frac{\log_2(\Delta/\epsilon)}{\lambda}$. Consider an information approach to the same problem. We choose to (or are forced to) have a particular partition size $\epsilon$. This will correspond to $A = \frac{\max(U) - \min(U)}{\epsilon}$. Our maximum possible uncertainty in bits is $\log_2 A$. It will take $n = \frac{\log_2 A}{h}$ steps into the future to add up to this uncertainty and beyond this our data stream is unpredictable.

We estimate $h$ using the limit of $h_L$ from Eq. \ref{eq:hL} in Eq. \ref{eq:h}, as discussed in Ref. \cite{cerbus2013} and elsewhere \cite{schurmann1996,crutchfield2003}. The undersampling bias in the $H(U^L)$ is corrected using Grassberger's method \cite{schurmann1996}, although this did not affect the value of $h$ very much. The $h_L$ typically reached $h$ at $L \simeq 10$.

\section{Excess entropy $E$}
\label{appendix:E}

While $h$ tells us about the unpredictability of $\overrightarrow{U^1}$ given $\overleftarrow{U}$, we may also want to know how much we actually learned about $\overrightarrow{U}$ from $\overleftarrow{U}$. This is the excess entropy $E$. It is in some sense the opposite of unpredictability. $E$ doesn't ask how much information we get from $\overrightarrow{U}$ upon measuring, but how much we don't get. We already know it. Stated mathematically \cite{schurmann1996,crutchfield2003}:
\begin{equation}
E = H(\overrightarrow{U}) - H(\overrightarrow{U} | \overleftarrow{U}) \equiv I(\overrightarrow{U} ; \overleftarrow{U})
\end{equation}
where  $I(\overrightarrow{U} ; \overleftarrow{U})$ is the mutual information shared between $\overrightarrow{U}$ and $\overleftarrow{U}$ \cite{cover1991}.

This $E$ is the information we got from $\overleftarrow{U}$ that reduces unpredictability. However, just like $h$, this is a statistical statement that doesn't tell us how to use that information. $E$ does provide us with a lower bound on the amount of information needed to make predictions, since we need to account for all correlations. No matter how it's done, $E$ bits will be necessary \cite{crutchfield2003}, otherwise we ignore some structure in the system.

An alternative expression is used to estimate $E$ \cite{crutchfield2003}:
\begin{equation}
E = \sum_{L = 1}^{\infty} (h_L - h)
\end{equation}
This calculation uses essentially the same quantities involved in estimating $h$. It turns out that for many chaotic systems, $h_L - h \propto 2^{-\gamma L}$ ($\gamma$ is some constant independent of $L$) \cite{crutchfield2003}. This empirical relationship has been shown to improve the estimation of $E$ \cite{crutchfield2003}. This expression will be used when possible.

\section{Crutchfield complexity $C$}
\label{appendix:C}

We now come to prediction using a statistical model. We must determine a set of special states called causal states $S$ \cite{crutchfield2012}. These will make up a minimal representation of our system for predictive purposes. In other words, we are trying to build the simplest possible statistical model of our data. For more details see Ref. \cite{shalizi2001}. There Shalizi $et$ $al.$ show that within the information theory framework, the approach described below is maximally predictive with a minimal amount of information needed.

A statistical model consists of a set of states and the transition probabilities between them. To determine $S$ consider all unique blocks of data $U^L$. One would like to make $L$ large to capture as many correlations as possible, but the finite amount of data means only finite $L$ can be statistically reliable. For our data, $L \simeq 10$ is a good compromise. This $L$ is also chosen because it is the value of $L$ at which $h_L$ typically reached $h$.

We now calculate the conditional probability $p(\overrightarrow{U}^L|\overleftarrow{u}^L)$ that any particular block $\overleftarrow{u}^L$ will give rise to any other block of the same length. If the conditional probability distributions conditioned on two blocks are the same, they are indistinguishable from a statistically predictive point of view. Thus block 1 and block 2 are equivalent, $u^L_1 \sim u^L_2$, if $p(\overrightarrow{U}^L|\overleftarrow{u}^L_1) = p(\overrightarrow{U}^L|\overleftarrow{u}^L_2)$. This process incorporates pattern recognition by construction, which is why $C$ was originally introduced as a complexity quantifier \cite{feldman1998,crutchfield2012}.

All equivalent blocks are then combined and organized into a set of predictive causal states $S$. For example, suppose there are only three states $u_1$, $u_2$, and $u_3$ (forget about $L$ here). If $p(\overrightarrow{U}|\overleftarrow{u}_1) = p(\overrightarrow{U}|\overleftarrow{u}_2) \ne p(\overrightarrow{U}|\overleftarrow{u}_3)$, then $u_1 \sim u_2 \nsim u_3$ and we have two causal states $s_1 = (u_1, u_2)$ and $s_2 = (u_3)$. Refer back to the example in Sec. \ref{sec:example}. It is apparent that if $P = 0.5$ (or 1) there is only one causal state, but if $P \neq 0.5$ (or 1), there are two causal states.

The Shannon information (entropy) contained in $S$ is the statistical complexity \cite{crutchfield2012,crutchfield1989}
\begin{equation}
C = H[S] = - \sum_{s} p(s) \log_2 p(s).
\label{eq:C}
\end{equation}
This is the total amount of information needed to statistically reproduce the data, as we shall soon see.

Here is how this prediction work in practice: we find the causal states $S$ as just described and so we also have the transition probabilities between the states $S$. Start out in some state $u$ belonging to a particular $s$. Determine the next $s'$ statistically using the known transition probabilities $p(s'|s)$ (the $'$ means the next step). Then determine a particular $u'$ belonging to this $s'$ according to $p(u'|s')$. This is symbolically represented by
\begin{equation*}
u \xrightarrow{u \in s} s \xrightarrow{p(s'|s)} s' \xrightarrow{p(u'|s')} u'.
\end{equation*}
Then repeat. In this way the data is reproduced in a statistical sense. In summary, we can write down the probability of any $u$ starting from any other $u$. This is statistical prediction.

We needed to know an amount of information $C = H[S]$ to carry out the above prediction program. That is, we need to ask (on average) $C$ ``yes" or ``no" questions in order to find the current state of the system, and then predict from there. By design, this connects with the system's predictability, since organizing the message's parts into causal states will affect the value of $C$.

We can appreciate the distinction between $C$ and $h$ by considering an unbiased coin flip. The system is maximally unpredictable with $h = 1$, since one has no clue as to what will come next. In contrast, $C = 0$ since no information is needed for statistical prediction. There is only one causal state. This may strike the readers as strange, since random data is supposedly impossible to predict. This is only true if we insist on a prediction that has absolute certainty. Here we are predicting statistically.

\begin{figure}[h!]
\hspace{-1.5em}
\includegraphics[scale = 0.35]{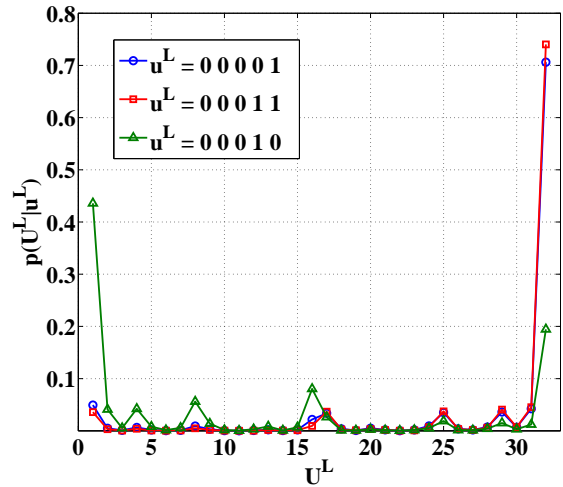}
\caption{An example of three conditional pdfs used to determine the causal states. The data used here is binarized turbulence data with $L = 5$ (giving a total of 32 possible states) and $Re = 3300$ ($\bigcirc = 00001$, $+ = 00011$, $\triangle = 00010$). The horizontal axis features all the possible future states while the vertical axis is the conditional probability that given a certain past state, any of those possible future states will occur. Here the distribution for states $\bigcirc$ and $\square$ appear similar while that for state $\triangle$ is quite different.}
\label{cond_pdf}
\end{figure}

When actually handling real data to identify $S$, one must deal with imperfections. These may be due to external noise or the finiteness of the amount of data. Regardless of the origin, one must set some sensible threshold to determine if two conditional probability distributions are the same, since they will never be identical. An example of some conditional probability distributions is shown in Fig. \ref{cond_pdf}. Two of the distributions are similar, indicating that the two states belong to the same causal state. The third distribution is entirely different. The task is to choose a sensible metric to make this distinction objectively.

We wrote a MATLAB program that uses a $\chi^2$ test to compare conditional probability distributions \cite{lehmann2005}. We use a 0.95 confidence level, but the results are not sensitive to this choice. Results from our method are in good agreement with another frequently used algorithm \cite{shalizi2003,shalizi_website}. In the end, of course, the choice has an element of subjectivity to it.

Note that alternative expressions for $h$ and $E$ are \cite{crutchfield2012,crutchfield2009}
\begin{equation}
h = H[\overrightarrow{U^1} | \overleftarrow{S}]
\label{eq:hS}
\end{equation}
and
\begin{equation}
E = I[\overrightarrow{S} ; \overleftarrow{S}] =  H[\overrightarrow{S}] - H[\overrightarrow{S} ; \overleftarrow{S}] = C - H[\overrightarrow{S} ; \overleftarrow{S}].
\label{eq:ES}
\end{equation}
Equations \ref{eq:hS} and \ref{eq:ES} say that the causal states serve as a sufficient representation. Equation \ref{eq:hS} also serves as a check on our determination of $S$ by comparing $h$ calculated with Eq. \ref{eq:hS} with our previous method from Eqs. \ref{eq:hL} and \ref{eq:h}. From Eq. \ref{eq:ES} we see that $C$ may be different from $E$. Actually, it can be shown that $C \ge E$. The difference between these two has various interpretations.

The interpretation of Crutchfield and coworkers is that a system may have some ``hidden" information, or crypticity $\chi = C - E$ \cite{crutchfield2009,mahoney2011}. Despite looking at the infinite $\overleftarrow{U}$, we missed out on the need to have an extra amount of information $\chi$ for prediction. Wiesner and coworkers have interpreted $\chi$ as the information erased at each step in the system's evolution \cite{wiesner2012}. If we were to simulate this process on a computer, $k_B T \chi$ (where $k_B$ = Boltzmann's constant and $T$ is the computer's temperature) would be the minimum thermodynamic cost. This is an extension of Landauer's work on computation. He was the first to suggest that the erasure of information has a thermodynamic cost \cite{landauer1996}.

\end{appendix}


\begin{thebibliography}{99}

\bibitem{tritton1988} D. J. Tritton, {\it Physical Fluid Dynamics} (Oxford University Press, USA, 1988)

\bibitem{tennekes1972} H. Tennekes, J. L. Lumley {\it A First Course in Turbulence} (MIT Press, Cambridge, Mass., 1972)

\bibitem{shannon1964} C. E. Shannon, W. Weaver, {\it The Mathematical Theory of Communication} (University of Illinois Press, Urbana, 1964)

\bibitem{cover1991} T. M. Cover, J. A. Thomas, {\it Elements of Information Theory} (Wiley, New York, 1991)

\bibitem{brillouin1962} L. Brillouin, {\it Science and Information Theory} (Academic Press, New York, 1962)

\bibitem{gershenfeld2000} N. Gershenfeld, {\it The Physics of Information Technology} (Cambridge University Press, Cambridge, 2000) 

\bibitem{crutchfield2012} J. P. Crutchfield, Nat. Phys. {\bf 8}, 17 (2012)

\bibitem{baker1996} G. L. Baker, J. P. Gollub, {\it Chaotic Dynamics: An Introduction} (Cambridge U. Press, 2nd Ed., Cambridge, 1996)

\bibitem{davidson2004} P. A. Davidson,  {\it Turbulence: An Introduction for Scientists and Engineers} (Oxford University Press, Oxford, 2004)

\bibitem{crutchfield2003} J. P. Crutchfield, D. P. Feldman, Chaos {\bf 13}, 25 (2003)

\bibitem{schurmann1996} T. Sch\"{u}rmann, P. Grassberger, Chaos {\bf 6}, 414 (1996)

\bibitem{ellison2009} C. J. Ellison, J. R. Mahoney, J. P. Crutchfield, J. Stat. Phys. {\bf 136}, 1005 (2009)

\bibitem{shalizi2001} C. R. Shalizi, J. P. Crutchfield, J. Stat. Phys. {\bf 104}, 817 (2001)

\bibitem{aurell1996} E. Aurell, G. Boffetta, A. Crisanti, G. Paladin, A. Vulpiani, Phys. Rev. E {\bf 53}, 2337 (1996)

\bibitem{leith1972} C. E. Leith, R. H. Kraichnan, J. Atmos. Sci. {\bf 29}, 1041 (1972)

\bibitem{kolmogorov1941} A. N. Kolmogorov, ``The local structure of turbulence in incompressible viscous fluids for very large Reynolds numbers," Dokl. Akad. Nauk. SSSR {\bf 30}, 299 (1941) (Proc. R. Soc. Lond. A {\bf 434} (reprinted))

\bibitem{kraichnan1994} R. H. Kraichnan, Phys. Rev. Lett. {\bf 72}, 1016 (1994)

\bibitem{shraiman2000} B. I. Shraiman, E. D. Siggia, Nature {\bf 405}, 639 (2000)

\bibitem{falkovich2001} G. Falkovich, K. Gawedzki, M. Vergassola, Rev. Mod. Phys. {\bf 73}, 913 (2001)

\bibitem{boeing} Boeing, http://www.boeing.com/boeing/commercial/...
...747family/pf/pf$\_$400$\_$prod.page,
4 July, 2014

\bibitem{mcminn1997}  J. D. McMinn, AIAA Guidance Navigation and Control Conference, AIAA–97–3532, (1997)

\bibitem{kellay2002} H. Kellay, W. I. Goldburg, Rep. Prog. Phys. {\bf 65}, 845-894 (2002)

\bibitem{cerbus2013} R. T. Cerbus, W. I. Goldburg, Phys. Rev. E {\bf 88}, 053012 (2013)

\bibitem{quax2013} R. Quax, A. Appolloni, P. M. A. Sloot, Eur. Phys. J. Spec. Top. {\bf 222}, 1389 (2013)

\bibitem{codling2008} E. A. Codling, M. J. Plank, S. Benhamou, J. R. Soc. Interface {\bf 5}, 813 (2008)

\bibitem{crutchfield2009} J. P. Crutchfield, C. J. Ellison, J. R. Mahoney, Phys. Rev. Lett. {\bf 103}, 094101 (2009)

\bibitem{boffetta2012} G. Boffetta, R. Ecke, Ann. Rev. Fluid Mech. {\bf 44}, 427 (2012)

\bibitem{kellay2012} H. Kellay, T. Tran, W. I. Goldburg, N. Goldenfeld, G. Gioia, P. Chakraborty, Phys. Rev. Lett. {\bf 109}, 254502 (2012)

\bibitem{landau1987} L. D. Landau, E. M. Lifshitz, {\it Fluid Mechanics} (Butterworth-Heinemann, 2nd Ed., Oxford, 1987)

\bibitem{schumacher2007} J. Schumacher, K. R. Sreenivasan, V. Yakhot, New. J. Phys. {\bf 9}, 89 (2007)

\bibitem{schumacher2014} J. Schumacher, J. D. Scheel, D. Krasnov, D. A. Donzis, V. Yakhot, K. R. Sreenivasan, PNAS {\bf 111}, 10961 (2014)

\bibitem{sreenivasan1997} K. R. Sreenivasan, R. A. Antonia, Ann. Rev. Fl. Mech. {\bf 29}, 435 (1997)

\bibitem{kadanoff2001} L. P. Kadanoff, Phys. Today {\bf 54}, 34 (2001)

\bibitem{brandstater1983} A. Brandst\"{a}ter, J. Swift, H. L. Swinney, A. Wolf, J. D. Farmer, E. Jen, J. P. Crutchfield, Phys. Rev. Lett. {\bf 51}, 1442 (1983)

\bibitem{ruelle1971} D. Ruelle, F. Takens, Commun. Math. Phys. {\bf 20}, 167 (1971)

\bibitem{swinney1978} H. L. Swinney, J. P. Gollub, Phys. Today {\bf 31}, 41 (1978)

\bibitem{wiesner2012} K. Wiesner, M. Gu, E. Rieper, V. Vedral, Proc. Roy. Soc. A {\bf 468}, 4058 (2012)

\bibitem{crutchfield1997} J. P. Crutchfield, D. P. Feldman, Phys. Rev. E {\bf 55}, R1239 (1997)

\bibitem{daw2002} C. S. Daw, C. E. A. Finney, E. R. Tracy, Rev. Sci. Instrum. {\bf 74}, 915 (2002)

\bibitem{palmer2000} A. J. Palmer, C. W. Fairall, W. A. Brewer, IEEE Trans. Geo. Remote Sensing {\bf 38}, 2056 (2000)

\bibitem{lehrman2001} M. Lehrman, A. B. Rechester, Phys. Rev. Lett. {\bf 87}, 164501 (2001)

\bibitem{feldman1998} D. P. Feldman, J. P. Crutchfield, Phys. Lett. A {\bf 238}, 244 (1998)

\bibitem{crutchfield1989} J. P. Crutchfield, K. Young, Phys. Rev. Lett. {\bf 63}, 105 (1989)

\bibitem{lehmann2005} E. L. Lehmann, J. P. Romano, {\it Testing Statistical Hypotheses, 3rd Ed.} (Springer, New York, 2005)

\bibitem{shalizi2003} C. R. Shalizi, K. L. Shalizi, J. P. Crutchfield, arXiv:cs/0210025v3 [cs.LG]

\bibitem{shalizi_website} C. R. Shalizi, K. L. Shalizi, {\it An Algorithm for Building Markov Models from Time Series}, http://vserver1.cscs.lsa.umich.edu/$\sim$crshalizi/CSSR/, 15 May 2013.

\bibitem{mahoney2011} J. R. Mahoney, C. J. Ellison, R. G. James, J. P. Crutchfield, Chaos {\bf 21}, 037112 (2011)

\bibitem{landauer1996} R. Landauer, Phys. Lett. A {\bf 217}, 188 (1996)

\end{thebibliography}
\end{document}